# Room-temperature chiral charge pumping in Dirac semimetals

Cheng Zhang[1,2], Enze Zhang[1,2], Weiyi Wang[1,2], Yanwen Liu[1,2], Zhi-Gang Chen[3*], Shiheng Lu[1,2], Sihang Liang[1,2], Junzhi Cao[1,2], Xiang Yuan[1,2], Lei Tang[1,2], Qian Li[1,2], Chao Zhou[1,2], Teng Gu[1,2], Yizheng Wu[1,2], Jin Zou[3,4*], Faxian Xiu[1,2*]

[1] State Key Laboratory of Surface Physics and Department of Physics, Fudan University, Shanghai 200433, China

[2] Collaborative Innovation Center of Advanced Microstructures, Nanjing 210093, China

[3] Materials Engineering, The University of Queensland, Brisbane QLD 4072, Australia

[4] Centre for Microscopy and Microanalysis, The University of Queensland, Brisbane QLD 4072, Australia

* Correspondence and requests for materials should be addressed to F. X. (E-mail: faxian@fudan.edu.cn), J. Z. (Email: j.zou@uq.edu.au), or Z-G. C. (Email: z.chen1@uq.edu.au)

**Abstract**

Chiral anomaly, a non-conservation of chiral charge pumped by the topological nontrivial gauge fields, has been predicted to exist in Weyl semimetals. However, until now, the experimental signature of this effect exclusively relies on the observation of negative longitudinal magnetoresistance at low temperatures. Here, we report the field-modulated chiral charge pumping process and valley diffusion in $Cd_3As_2$. Apart from the conventional negative magnetoresistance, we observe an unusual nonlocal response with negative field dependence up to room temperature, originating from the diffusion of valley polarization. Furthermore, a large magneto-optic Kerr effect generated by parallel electric and magnetic fields is detected. These new experimental approaches provide a quantitative analysis of the chiral anomaly phenomenon which is inaccessible previously. The ability to manipulate the valley polarization in topological semimetal at room temperature opens up a brand-new route towards understanding its fundamental properties and utilizing the chiral fermions.

**Introduction**

Conservation laws arising from preserved symmetries are the base of modern physics.[1] However, when a classically preserved symmetry is broken upon quantization, such conservation may encounter a so-called quantum anomaly.[2] The chiral anomaly, one of the typical examples, manifests itself in non-conservation of chiral charge induced by topological nontrivial gauge fields.[2, 3] As a long-pursued topic in quantum field theory context, Weyl fermions host definite chiral charges with opposite signs,[2] whose chirality is defined by the sign of its spin polarization along the momentum direction. Weyl fermions with different chiralities are independently coupled to external fields, resulting in a separate charge conservation.[2, 4] With the presence of parallel electric and magnetic fields, Weyl fermions possess a non-conserved chiral charge, *i.e.*, the chiral anomaly.

Despite its crucial role in the description of elementary particles, the study of Weyl fermions in condensed matter physics has only been paid intensive attentions in recent years.[5, 6, 7] The condensed matter analogy of Weyl fermions was proposed as the emergent quasiparticle excitations of certain novel gapless topological matter, which is named as Weyl semimetal.[4] In a Weyl semimetal, the band structure shows semimetal behavior with the conduction and valence bands intersecting only at some discrete points, denoted as Weyl nodes.[4, 6] A non-zero Chern number is expected to emerge when the Fermi surface encloses a Weyl node, which can be viewed as a Berry curvature singular point.[2, 4] With time-reversal and inversion symmetries preserved, the Dirac nodes can be formed by two degenerated Weyl nodes, driving the system into a Dirac semimetal, a close sibling of the Weyl semimetal.[8] In Dirac semimetals, these nodes with opposite chiralities are distinguished by the point-group index or isospin and will split under external magnetic field.[9, 10, 11] Typically, additional crystalline point-group symmetry is required to prevent the overlapping Weyl nodes with different chiralities from annihilation.[8, 9] As the Weyl nodes are topologically-protected objects with well-defined chirality, the Dirac/Weyl semimetals are predicted to harbor many exotic effects, such as surface Fermi arc states[5] and axion strings.[12] Among them, the chiral anomaly raises particular interests owing to its important role in the 4-dimensional quantum Hall boundary state.[13] And it also promises intriguing transport phenomena, such as negative magnetoresistance (MR)[14], anomalous Hall effect[15] and nonlocal valley transport[16], thus holding prospects in valleytronics.

In Weyl semimetals, electrons can be pumped from one node to the other due to the chiral anomaly. This chirality-dependent charge pumping rate can be characterized by,[6]

$$\frac{\partial \rho_\chi}{\partial t} = \chi \frac{e^2}{4\pi^2 \hbar^2} \mathbf{E} \cdot \mathbf{B}, \qquad (1)$$

where $\rho_\chi$ is the pumped charge, $e$ is the electron charge, $\mathbf{E}$ and $\mathbf{B}$ are the external electric and magnetic fields, respectively. Such a pumping process will be compensated by the depletion of valley charge, characterized by the inter-valley scattering time $\tau_v$. Eventually, the Weyl nodes with opposite chirality $\chi$ will acquire a different chemical potential as schematically illustrated in Fig. 1a. Here, we would like to point out that for Dirac semimetals, the Weyl nodes with opposite chirality are distinguished by isospin although they are degenerated in the **k**-space.[16, 17] And external magnetic fields may further induce the splitting of Weyl nodes.[9] Hence, the phenomenon of chiral anomaly remains in Dirac semimetals.

One important signature of the chiral anomaly is the large negative MR with unusual anisotropy with respect to the angle between **E** and **B**.[14, 18] The negative MR was firstly observed in $Bi_{1-x}Sb_x$ system at the critical state between topological insulator and trivial band insulator.[19] After several years exploring the existence of Weyl fermions in crystalline solids, $Na_3Bi$ and $Cd_3As_2$ have been theoretically predicted and experimentally confirmed as Dirac semimetals.[9, 10, 20, 21, 22, 23] Soon after that, Weyl semimetal states were discovered in the inversion-symmetry-breaking TaAs family.[24, 25] Since then, the negative longitudinal MR, reminiscent of $Bi_{1-x}Sb_x$, has been frequently observed in these materials at low temperatures,[26, 27, 28, 29, 30, 31] serving as the sole experimental evidence for the chiral anomaly so far.

Here, we demonstrate the existence of the chiral anomaly in $Cd_3As_2$ based on three independent experimental evidences: the negative longitudinal MR, the valley transport, and the **E·B**-generated magneto-optical Kerr effect (MOKE). Surprisingly, in stark contrast to the strong temperature dependence of negative MR, the chiral pumping process is robust against thermal perturbation and persists up to room temperature as revealed by the following valley transport and MOKE experiments. It is noted that such chirality-polarized states are directly coupled with spin, orbit, and valley degree of freedom,[6, 16] thus they may be further exploited for electronic devices.

**Results**

**Negative magnetoresistance.** To investigate the influence of chiral anomaly on transport, we fabricated Hall bar devices based on $Cd_3As_2$ nanoplates. These nanostructures were grown by chemical vapor deposition method (refer to Supplementary Fig. 1 and Supplementary Note 1 for sample charactization). The magnetotransport properties of $Cd_3As_2$ nanoplates were systematically studied in our previous work.[32] It has been demonstrated that $Cd_3As_2$ transforms into semiconductors when approaching two-dimensional limit. Here we specifically chose thick nanoplates (> 80 nm) to ensure their semimetallic nature. Figure 1b summarizes the angle-dependent MR of the $Cd_3As_2$ nanoplates at 2.5 K. As illustrated in the inset, **B** was rotated from in-plane to out-of-plane with $\theta$ defined as the angle between **B** and the applied current **I**. A large positive and linear MR along with a strong anisotropy respective to the field direction was obtained, similar to those observed in $Cd_3As_2$ bulk crystals[11]. Remarkably, by carefully investigating the MR behavior as **B** approaches **I**, we find a clear dip emerging in the MR curve, which vanishes rapidly even if $\theta$ changes only a few degrees (Fig. 1c). Owing to its large amplitude and the corresponding field range (-2~2 T), this dip is distinct from weak-localization or Shubnikov-de Haas (SdH) oscillations. For the magnetic field over 2 T, the MR turns upwards, very similar to that in $Bi_{1-x}Sb_x$ and TaAs.[19, 28] Supplementary Table 1 and Supplementary Note 2 give a detailed analysis on the origin of this transition from negative to positive MR.

To exclude the influence from the intrinsic crystal anisotropy, we designed three different geometries to conduct transport measurements as shown in Fig. 2a. The dip and the strongly suppressed MR trend disappear as long as **B** and **I** are not parallel (Fig.

2b); and the positive MR in geometry II and III is originated from both the vanishing of the chiral anomaly and the restore of orbital MR due to the Lorentz force. These control experiments strongly suggest that the detected negative MR originates from the interplay between the electric and magnetic fields instead of the anisotropy from the material itself, thus presenting a consolidate evidence of the chiral magnetic effect in the presence of $\mathbf{E}\cdot\mathbf{B}$ fields. The reason for the observation of negative MR in $Cd_3As_2$ nanoplates is the relatively low carrier density and the corresponding low Fermi level in the as-grown nanoplates (refer to Supplementary Figs 2-5 and Supplementary Note 2 for the magnetotransport analysis and carrier density information). This chiral magnetic effect becomes stronger in transport as the Fermi level approaches the Weyl nodes.[14] Theoretically, it can be understood that in the quantum limit where only the lowest Landau level is occupied, the axial current induced by chiral anomaly is relaxed with an inter-valley scattering rate and a linear dependence of positive conductivity can occur.[4, 33] Moreover, the theoretical study has been extended to the semi-classical regime with conventional Boltzmann equation applied.[14] In the weak magnetic field regime where multiple Landau levels are occupied, this anomaly-induced magneto-conductance correction is proportional to $\mathbf{B}^2$.[14] As shown in Fig. 2c, the positive magneto-conductance can be well fitted by quadratic function $\sigma_B-\sigma_0=\alpha\mathbf{B}^2$ below 1T. Since the longitudinal MR is a mixture of axial current and original background, it is not appropriate to calculate the inter-valley scattering rate as the theoretical formula describes a pure conductance correction from chiral anomaly. The experimental coefficient α emerges below 150 K and becomes saturated below 50 K. This strong temperature and field dependence is consistent with the recent observation in ZrTe$_5$[29], Na$_3$Bi[20], Cd$_3$As$_2$ nanowires[27, 31], and TaAs[28, 30], possibly due to the competition between the axial current and the conventional linear or quadratic MR background. At low fields, the anomaly-related contribution dominates the MR while it gradually saturates at high fields.[14] Also, the transport life time and related momentum relaxation, both of which can be influenced by thermal fluctuation, will affect the overall MR behavior.

**Nonlocal transport.** However, recent studies have raised concerns about the negative MR alone as a reliable evidence for the chiral anomaly.[30, 34, 35] One of the major challenges is how to explicitly distinguish the axial-current-induced negative MR from the conventional MR anisotropy due to crystal anisotropy or inhomogeneous current distribution.[30, 35] To overcome this problem, Parameswaran *et al.*[16] proposed a nonlocal experiment that can isolate the axial current contribution and conventional MR in real space through valley diffusion. With a non-zero $\mathbf{E} \cdot \mathbf{B}$ term, the charge imbalance between Weyl nodes can induce valley polarization and the corresponding polarized states will diffuse in real space like spin. Because the relaxation process involves large quasi-momentum transfer or scattering between different point-group representations, these valley polarization will experience a slow relaxation characterized by the inter-valley scattering length $L_v$.[16] Here, the degree of freedom of valley acts as a pseudospin index that can be effectively modulated by the electromagnetic field.[16] If a detecting magnetic field is applied in the valley diffusion regime, the valley polarization can be converted into a nonlocal voltage (Fig. 3a), taking the same principle as the inverse spin Hall effect.[36] This nonlocal voltage follows a diffusion formula,[16]

$$|V_{\text{NL}}(x)| \propto V_{\text{SD}} e^{-L/L_{\text{v}}}, \qquad (2)$$

where $V_{\text{SD}}$ is the applied source-drain voltage, $L$ is the length between the Hall bar terminals as illustrated in Fig. 3a. From this equation, we find that the nonlocal response can survive a long distance if the inter-valley scattering is weak.

To detect such a nonlocal response, firstly we designed a specific device geometry as schematically illustrated in Fig. 3b. We notice that a nonlocal voltage can also be induced by the stray charge current in classical diffusive transport but it should possess the same field dependence with $R_{12}$. A mixed contribution from both the Ohmic nonlocal resistance and the valley diffusion is expected, since both of them follow the exponential damping rule with diffusion length $L$.[37, 38] But the contribution of stray charge current can be controlled by changing the ratio of $L/W$ according to the van der Pauw formula[37]. Following this idea, we made a device with different lateral channel width by focused ion beam (FIB) etching as shown in Fig. 3c. In this way, the nonlocal voltage due to chiral anomaly can be distinguished from the conventional Ohmic diffusion. By sweeping **B** which is parallel to the applied current, we measured the **B**-dependence of the two-terminal local resistance $R_{\text{L}}$ (resistance of terminal 1-2, $R_{12}$) and nonlocal resistance $R_{\text{NL}}$ ($R_{34}$ and $R_{56}$). As shown in Fig. 3d, a negative MR was firstly observed and followed by an upturn after 5 T, similar to what we have shown in Fig. 1c. In the meantime, both $R_{34}$ and $R_{56}$ acquire a negative field dependence from 0 to 9 T as shown in Fig. 3e. Owing to the fact that the nonlocal response from the valley diffusion vanishes at zero magnetic field, the Ohmic contribution can be evaluated and hence subtracted from the nonlocal resistance. Fig. 3f is the extracted pure nonlocal resistance. To verify if the detected nonlocal resistance ($R_{34}$ and $R_{56}$) is partially originated from valley diffusion, we plot the resistance ratio between two different channels ($R_{56}/R_{34}$ and $R_{56\text{-NL}}/R_{34\text{-NL}}$) as a function of **B**. The ratio of $R_{56}/R_{34}$ is close to 0.25 around zero field at different temperatures (Fig. 3g). The dash line marked as "Ohmic" is the theoretical value for $R_{56}/R_{34}$ from van der Pauw formula by considering the device geometry. The real geometry of conducting channel is hard to determine precisely since the electrodes are a bit wide comparing with the channel, which should be the origin of the small deviation from the "Ohmic" dash line. Increasing magnetic field will strongly suppress the ratio to nearly zero (Fig. 3g). The field dependences of $R_{34}$ (Fig. 3e) and $R_{56}/R_{34}$ (Fig. 3g) cannot be simply explained by the conventional Ohmic diffusion effect. Theoretically, the Ohmic diffusion is determined by van der Pauw formula derived from the Poisson equation, which is unlike to be influenced so strongly by magnetic field. Especially the suppression of the ratio of $R_{56}/R_{34}$ at high field is anomalous since terminals 3-4 and 5-6 are actually symmetric with only different width (which should only give an extra constant ratio based on Ohmic diffusion). Meanwhile, the observation of negative MR in both sides also excludes the influence of the current jetting effect (refer to Supplementary Note 3 for detailed analysis). On the other hand, if we consider the valley diffusion as part of the nonlocal resistance, the suppression of $R_{56}/R_{34}$ ratio will be understandable. The nonlocal voltage by valley diffusion is negative and only determined by the diffusion length and the voltage on the diffusion channel. Compared with valley diffusion, the Ohmic loss decays fast with the decrease of diffusion channel width. Here, $x_2/x_1$ is 0.5 and the diffusion length $L$ is the same for 3-4 and 5-6 terminals. Therefore, the theoretical value of $R_{56\text{-NL}}/R_{34\text{-NL}}$ should be 0.50, marked as "valley" dash line. We can see from Fig. 3g that the experimental ratio of $R_{56\text{-NL}}/R_{34\text{-NL}}$ is quite close to the theoretical "valley" dash line. Meanwhile, this assumption also explains the strong decrease of $R_{56}/R_{34}$ at high field. The original nonlocal resistance has two contributions from chiral anomaly $R_{\text{valley}}$ and stray charge currents $R_{\text{Ohmic}}$. For the Ohmic part, the ratio should be field-independent. The valley contribution is zero at zero field and will increase with magnetic field. Note that $R_{\text{valley}}$ is negative relative to $R_{\text{Ohmic}}$. $R_{56\text{-NL}}/R_{34\text{-NL}}$ is around 0.5, larger than $R_{56}/R_{34}$ at zero field. Therefore, with the increase of magnetic field, the absolute value of $R_{\text{valley}}$ increases and the total nonlocal resistance ratio $R_{56}/R_{34}$ decreases. A small deviation of $R_{56\text{-NL}}/R_{34\text{-NL}}$ from 0.5 is also witnessed, mainly from the different contact conductance ($\delta_{34} > \delta_{56}$, consistent with the device picture in Fig. 3c). The combination of analysis on $R_{56}/R_{34}$ and $R_{56\text{-NL}}/R_{34\text{-NL}}$ strongly suggests that the detected nonlocal resistance consists of both contributions from valley and Ohmic diffusion.

Another important aspect of valley transport is the length dependence, from which an important parameter, inter-valley scattering length $L_v$, can be determined. Here, we employed three pairs of well-aligned "nonlocal" Hall bars ($R_{34}$, $R_{56}$, and $R_{78}$ in Fig. 4a) to measure the nonlocal resistance with different diffusion length. Similarly, the nonlocal resistance $R_{\text{NL}}$ from all the nonlocal terminals ($R_{34}$, $R_{56}$, and $R_{78}$) adopts a complete opposite dependence on **B** in comparison with the local resistance $R_{12}$. After extracting the Ohmic contribution (Supplementary Figs 6-8 and Supplementary Notes 4-5), the nonlocal resistance of valley transport at 100 K is shown in Fig. 4b with the corresponding local resistance displayed in the inset. As expected, a strong $R_{NL}$ reduction is observed with the longer lateral distance.

To perform the quantitative analysis, a dimensionless coefficient $\alpha_{NL}$ is introduced as the strength of the nonlocal response, defined as $\alpha_{NL} = R_{NL}/R_L$. In analogy to the inverse spin Hall effect, $\alpha_{NL}$ arising from the valley diffusion is given by [16]

$$\alpha_{NL} = \frac{R_{NL}}{R_L} = -\left(\frac{B}{\delta+B}\right)^2 e^{-\frac{L}{L_v}}, \tag{3}$$

where $\delta$ is proportional to the conductance at the metal contact regime. It is evident that $R_L$ and $R_{NL}$ hold opposite sign of the field dependence (also, in agreement with the experimental data in Fig. 3f and Fig. 4b). Figure 4c shows $\alpha_{NL}$-**B** curves, which adopt a parabolic dependence on the magnetic field. And they can be well fitted by Equation (3), yielding a value of $L_v$ around 1.5 μm with small variation with temperature (Supplementary Fig. 9 and Supplementary Note 6). A tendency of saturation at high fields develops at 100 K. Such phenomenon may originate from the so-called quantum limit effect that the nonlocal response is limited by the metal contact instead of the relaxation at the sample.[16] Figure 4d is a two-dimensional plot of $\alpha_{NL}$ as a function of temperature and magnetic field. At high magnetic field, $|\alpha_{NL}|$ first increases with lowering the temperatures due to an enhanced valley diffusion process by reducing thermal fluctuation, then it becomes saturated and even decreases a little due to the quantum limit effect. Since $L_v$ shows weak temperature dependence, the nonlocal ratio $\alpha_{NL}$ is mainly affected by the metal contact conductance when changing the temperature. We further estimated the contact resistance through the comparison of two-terminal and four-terminal measurements. The temperature dependence of contact resistance (Supplementary Fig. 10) agrees with the trend of nonlocal ratio shown in Fig. 4d. Here, the nonlocal signals persist up to 300 K, reflecting the robustness of chiral anomaly effect.

A more accurate way of measuring $L_v$ is to detect the length scaling of nonlocal signals by performing a line fit to the semi-log plot of $\alpha_{NL}$ against the lateral length $L$. As shown in Fig. 5, the nonlocal ration $\alpha_{NL}$ decreases exponentially with $L$. By extrapolating the dependence of $\alpha_{NL}$ on $L$ in Fig. 5, a high nonlocal ratio over 10% is expected for $L < 1$ μm. A relatively long $L_v$ of ~ 2 μm is extracted (Fig. 5 inset), similar to the value obtained from the field dependence (~ 1.5 μm). This high nonlocal ratio in the low-carrier-density samples actually reveals a strong response in the mesoscopic charge distribution to the changes in chirality states. A comparison between samples with different Fermi levels (Supplementary Figs 11-12, Supplementary Table 2 and Supplementary Note 7) reveals that the chiral anomaly effect will largely suppressed when the Fermi level moves away from the

**Kerr effect.** Besides the influence on the electric transport, the cooperation of **E** and **B** fields may also affect the optical properties. Since the field modulation on the Weyl nodes is intimately connected to Berry curvature and spin texture, it may have an impact on the optical activities. The Kerr effect has been widely used as a precise probe for dielectric tensor [39] and magnetization [40] by detecting the polarization change of reflected light. Here, we carried out rotational magneto-optical Kerr effect (ROT-MOKE) measurements on the bulk $Cd_3As_2$ single crystals (refer to Supplementary Fig. 13 and Supplementary Note 8 for crystal charactization) as illustrated in Fig. 6a. Initially, no Kerr signal from the crystal surface was observed when a magnetic field (up to 2000 Oe) was rotated in-plane, as one can expect from a system with preserved inversion and time-reversal symmetries. However, when a DC electric bias was applied on the two ends of the samples, the large Kerr rotation signals were dramatically produced with the increase of the current density (0~144 mA mm$^{-2}$), as demonstrated in Fig. 6b. Figure 6c shows the ROT-MOKE data under different magnetic fields with a fixed current density of 144 mA mm$^{-2}$. The ROT-MOKE curves adopt a cosine-function dependence on the angle $\theta$ between the electric and magnetic fields. The absolute value of Kerr rotation reaches a maximum when the electric and magnetic fields are parallel ($\theta = 0$°), and in turn it becomes zero when they are perpendicular ($\theta = 90°$). In addition, we performed the ROT-MOKE experiments with the magnetic field in the x-z plane (Supplementary Fig. 14 and Supplementary Note 9). When the electric field is parallel to the **B**-rotating plane, the similar 360°-period ROT-MOKE curves were observed (Supplementary Fig. 14b). However, when the bias was added perpendicularly to the **B**-rotating plane, no Kerr signal was ever detected (Supplementary Fig. 14d) with the vanishing of $\mathbf{E} \cdot \mathbf{B}$. Supplementary Figs 15-17 show detailed relationship of Kerr rotation with magnetic field, current density, and temperature. Similar with the nonlocal transport, the MOKE signals also show a weak temperature dependence (Supplementary Fig. 17).

**Discussion**

It is of vital importance to investigate the possible underlying mechanism for this novel Kerr effect. Previous theoretical work[41] suggests that Weyl semimetal with broken time-reversal symmetry is expected to induce Kerr/Faraday rotation. However, our experiments do not fit into this scenario since only **B** alone cannot lead to the Kerr effect as shown in Fig. 6b. Similarly, it is also different from the conventional electric-optical Kerr effect, which is usually induced by electrical field alone.[42]

One possible origin is the variable optical gyrotropic coefficient that is modulated by $\mathbf{E}\cdot\mathbf{B}$ fields through the chiral anomaly effect. A study by Hosur and Qi shows that the charge imbalance between the Weyl nodes leads to a non-zero gyrotropic coefficient $\gamma$, a Hall-like contribution to the dielectric tensor.[43] For a single Weyl node, $\gamma$ is given by $\gamma(\omega) = \mathrm{i}\frac{\chi\mu_\chi e^2 \tau_{\text{intra}}}{6\pi^2 \varepsilon_0 \hbar^2 \omega}$ with $\omega$ being the light frequency and $\tau_{\text{intra}}$ being the intra-valley scattering time.[43] They focused on the low frequency case of the detection light to ensure an equilibrium state of the Weyl nodes under $\mathbf{E}\cdot\mathbf{B}$ fields. Thus only circular dichroism is expected. However, in our case, the laser frequency is in the visible light range. If the light frequency has exceeded the low frequency limit (*i.e.*, $\omega\tau_{\text{intra}} \ll 1$ is not satisfied), the Weyl nodes will be in non-equilibrium configurations where Hall conductivity of doped Weyl nodes becomes complex due to the finite frequency electromagnetic response (refer to Supplementary Note 10 for details). In this case, the effective time-reversal symmetry will be broken and that may lead to the Kerr effect. Since $L_v$ has been obtained by the nonlocal experiments, we can calculate the inter-valley scattering time $\tau_v$ by the diffusion formula $L_v = \sqrt{D\tau_v}$. Here, $D = \mu k_B T/e$ is the charge diffusion coefficient, where $\mu$ is electron mobility and $k_B$ is the Boltzmann constant. Hence, $\tau_v$ is determined to be $2 \times 10^{-10}$ s. On the other hand, we may use the transport life time at room temperature ($2 \times 10^{-13}$ s) deduced from the Drude model as an estimation of the intra-valley momentum relaxation rate ($\tau_{\text{intra}}$). So we can in turn verify the initial assumption on the occurrence of Kerr effect (finite frequency electromagnetic response from the charge pumping process). Note that these two relaxation rates are calculated based on the data of the $Cd_3As_2$ nanoplates. In single crystals, the corresponding values are expected at least one order of magnitude larger owing to the much higher mobility[11, 44] and the absence of boundary scattering. Here, the laser frequency $\omega$ is $2.8 \times 10^{15}$ rad s$^{-1}$. $\omega\tau_v \gg 1$ ensures the independence of charge pumping in each Weyl node. And $\omega\tau_{\text{intra}} \gg 1$ indicates that the Weyl nodes are in fact not able to reach equilibrium within the light frequency. Thus a finite frequency electromagnetic response is anticipated with a complex Hall conductivity of gyrotropic tensor and the effective time-reversal symmetry breaking, leading to the emergence of the Kerr effect.[45]

Another possible origin for the Kerr effect is the spin-polarization induced by the $\mathbf{E} \cdot \mathbf{B}$ fields. The chiral anomaly leads to the polarization of Weyl nodes. The charge transfer between a pair of Weyl nodes is accomplished through the surface Fermi arcs. As demonstrated by recent photoemission studies, the Fermi arcs are highly spin polarized.[46, 47] Despite the relatively large Kerr signal, the observed angle and magnetic field dependence of Kerr rotation fits to the scenario of spin-Kerr effect from current induced spin polarization.[48] Either way, the dependence of Kerr effect on $\mathbf{E} \cdot \mathbf{B}$ field suggests its intimate relationship with the chiral anomaly. So far, this novel Kerr effect is still not fully understood and thus deserves further theoretical and experimental investigations.

In conclusion, our work presents several evidences for the chiral anomaly in Dirac semimetal $Cd_3As_2$ and further demonstrates the field-modulated valley transport. We find that the nonlocal response can serve as an accurate way to measure the inter-valley relaxation rate. The ability to manipulate the chiral charge at room temperature is a significant step towards quantum electronics. The robustness of the chiral anomaly not only reflects the topological character of the Weyl nodes, but it also reveals the prospects of utilizing chiral fermions in the future low-dissipation valleytronic applications.

**Methods**

**$Cd_3As_2$ nanostructure growth**

The $Cd_3As_2$ nanoplates were grown using $Cd_3As_2$ powders as the precursor in a horizontal tube furnace, in which argon was a carrier gas. Prior to the growth, the furnace was pumped and flushed with argon several times to remove water and oxygen. The temperature was ramped to the growth temperature within 15 min, held constantly for 20 min, and then was cooled down naturally over ~2 h in a constant flow of argon before the substrates were removed at room-temperature. The precursor boat was placed in the hot center of the furnace (held at 720 °C), while the smooth quartz substrates were placed in the down-stream within a very small temperature range from 150 °C to 200 °C. The smooth quartz substrates then appeared shining to the naked eyes.

**Material characterizations**

The crystal structures of the synthesized products were characterized by X-ray diffraction (XRD), recorded on an X-ray diffractometer equipped with graphite monochromatized, Cu Kα radiation (λ = 1.5418 Å). The morphological, structural, and chemical characteristics of the synthesized products were investigated by scanning electron microscopy (SEM, JEOL 7800 and 7001) and transmission electron microscopy (TEM, FEI F20) equipped with energy-dispersive X-ray spectroscopy (EDS).

**Device fabrication and measurement**

The $Cd_3As_2$ Hall bar devices were fabricated by electron beam lithography technique and plasma-etched by argon for 2 min at the electrode regime before the deposition of Cr/Au (5nm/150nm) electrodes. The FIB process is achieved using a low current of 50 pA, 30 kV to avoid strong damage to the sample. The magnetotransport measurements of the devices were carried out in a Physical Property Measurement System (Quantum design) with constant DC current (μA range) through the devices.

**$Cd_3As_2$ bulk crystal growth**

High-quality $Cd_3As_2$ single crystals were synthesized by self-flux growth method in a tube furnace with stoichiometric amounts of high-purity Cd powder (4N) and As powder (5N). Mixed elements were sealed in an alumina crucible inside an iron crucible under argon atmosphere. The iron crucible was heated to 800-900°C and kept for 24 hours, then slowly cooled down to 450°C at 6°C per hour. Then the crucible was kept at 450°C for more than one day before cooled naturally to room temperature. The superfluous Cd flux was removed by centrifuging in a vacuum quartz tube at 450°C. To avoid possible surface contaminations and oxidation, the bulk crystal was measured within three days after exposed to the atmosphere. Supplementary Fig. 18 and Supplementary Note 11 present the transport properties of the as-grown bulk crystal.

**MOKE experimental setup**

A 2-dimensional vector rotate magnet was used to generate a rotating magnetic field in the y-z plane, which is parallel to the sample surface as illustrated in Fig. 1b. The incident laser was *p*-polarized with a wavelength of 670 nm. In the MOKE measurements, the Kerr rotation angle from the *p*-polarized light can be quantified as a function of field strength and orientations.[40] The measurement of Kerr signal was achieved by nearly-crossed polarizers with two-degree misalignment for light detection. The incident angle of laser is around 45°. The MOKE experiments were performed at room temperature with a square-shape-like $Cd_3As_2$ bulk flake with thickness around 100 μm. A constant DC electric bias was applied across the sample. The electrical bias applied on the sample was in the range of 0-500 mV, mainly dropping on the contact regime. The actual voltage drops on the sample was less than 5 mV with 115 mA current applied on the $Cd_3As_2$ bulk crystal. The Kerr signal was re-checked by switching the direction of the applied current.

## Data availability

The data that support the findings of this study are available from the corresponding author on request.

## Acknowledgements

F. X. acknowledges valuable discussions with P. Hosur, X-L. Qi, D. Pesin and N. Mandal. This work was supported by the National Young 1000 Talent Plan and National Natural Science Foundation of China (61322407, 11474058, 61674040). Part of the sample fabrication was performed at Fudan Nano-fabrication Laboratory. This work was also supported by Australian Research Council. Z-G. C. thanks QLD government for a smart state future fellowship. The Australian Microscopy & Microanalysis Research Facility and the Queensland node of the Australian National Fabrication Facility are acknowledged for providing characterization facilities. C. Z. thanks Junxue Li and Jianhui Liang for helps in the MOKE experiments.

## Author contributions

F.X. conceived the ideas and supervised the overall research. J.C. and S.L. synthetized $Cd_3As_2$ single crystal. C.Z., Q.L., C.Z., T.G. and Y.W. performed the MOKE measurements. Z-G.C. S.L. and J.Z. synthetized $Cd_3As_2$ nanostructures and performed crystal structural analysis. E.Z. C.Z. and W.W. fabricated the $Cd_3As_2$ Hall bar devices. C.Z., Y.L. and X.Y. performed the electrical transport measurements. C.Z. analyzed the transport data. Y.W. and L.T. provided useful suggestions. C.Z. and F.X. wrote the paper with helps from all other co-authors.

## Competing financial interests

The authors declare no competing financial interests.

**Figure captions**

**Figure 1| Illustration of the charge pumping process and the angle-dependent MR in the Cd$_3$As$_2$ nanoplates.** (**a**) Charge pumped from one Weyl node to the other in the presence of chiral gauge fields. This pumping process is also applied to Dirac semimetals, whose nodes are degenerated but distinguished by isospin. (**b**) The angle-dependent MR curves of sample N1. The inset is the schematic view of the transport measurement setup, showing **B** rotating in the x-z plane. Here, $\theta$ is defined as the angle between **B** and **I**. For a large $\theta$, the sample exhibits a positive **B**-linear MR. (**c**) An enlarged view of MR of sample N1 as $\theta$ approaches zero. MR decreases rapidly as $\theta$ approaches zero. When $\theta$ is near zero, a clear dip was observed in the -2~2 T range, followed by an upturn of MR in the larger fields.

**Figure 2| Chiral anomaly induced negative longitudinal MR in sample N1.** (**a**) MR in three different geometries in transport measurements at 2.5 K. (**b**) The negative MR only appears when **E** is parallel with **B**. The control experiments with geometries II and III exclude the crystal anisotropy effect on the observed negative MR. (**c**) The magneto-conductance change $\sigma_B - \sigma_0$ at different temperatures, following a quadratic dependence, and it decreases after a critical field. The inset summarizes the fitting coefficient $\alpha$ at different temperatures.

**Figure 3| Detection of valley transport in sample N2.** (**a**) Schematic view of the valley diffusion process. Parallel (antiparallel) electric and magnetic fields generate the charge imbalance between two Weyl nodes due to the chiral anomaly. The charge imbalance of different valleys can diffuse across the sample and can be converted into a nonlocal voltage along the applied magnetic field direction. (**b**) Schematic view of the nonlocal resistance measurement with different diffusion channel width. Current is applied through terminal 1-2 while terminals 3-4 and 5-6 are used to measure the nonlocal resistance. The diffusion channel width ($x_1$ and $x_2$) for 3-4 and 5-6 are 2 μm and 1 μm, respectively. The diffusion length $L$ is 2 μm. (**c**) The scanning electron microscopy picture of the Cd$_3$As$_2$ device. The white scale bar corresponds to 2 μm. The contact regime in terminals 3-4 is slightly larger than that of 5-6. (**d**) The two-terminal local resistance ($R_{12}$) of the Cd$_3$As$_2$ device at 20 K. (**e**) The nonlocal resistance ($R_{34}$ and $R_{56}$) at 20 K. (**f**) The pure nonlocal resistance ($R_{34\text{-NL}}$ and $R_{56\text{-NL}}$) after subtracting the Ohmic diffusion at 20 K. (**g**) Resistance ratio of $R_{34}/R_{56}$ and $R_{34\text{-}}$

$_{NL}/R_{56\text{-NL}}$ as a function of *B* at different temperatures. The dash line "valley" and "Ohmic" correspond to 0.50 and 0.17, respectively. $R_{34\text{-NL}}/R_{56\text{-NL}}$ at low field (<4 T) is not plotted since the value of $R_{34\text{-NL}}$ or $R_{56\text{-NL}}$ is close to zero and the corresponding ratio is easily affected by fluctuations or small curvatures.

**Figure 4| Diffusion length dependence of valley transport.** (**a**) Schematic view of the nonlocal resistance measurement in a Hall-bar-geometry device with one pair of local terminal (1-2) and evenly spaced three pairs of nonlocal terminals (3-4, 5-6, 7-8). The distance between terminal 3 and 4 (or 5 and 6, 7 and 8) is about 2 μm. The diffusion lengths of terminals 3-4, 5-6, and 7-8 are 2.5 μm, 5.0 μm and 7.5 μm, respectively. (**b**) The extracted nonlocal resistance $R_{NL}$ ($R_{34}$, $R_{56}$, and $R_{78}$) of sample N1 measured in the Hall bar geometry at 100 K. The inset is $R_L$-**B** curves from terminal 1-2. (**c**) The nonlocal ratio $\alpha_{NL}$-**B** of sample N1 and the corresponding fitting curves at different temperatures. (**d**) A two-dimensional plot of $\alpha_{NL}$ in sample N1 as a function of temperature and magnetic field.

**Figure 5| Length dependence and calculated valley-relaxation length of the valley transport.** The calculated $|\alpha_{NL}|$ as a function of diffusion length *L* in sample N1. It decays exponentially with the increasing of lateral length *L*. The light blue background is a guide line to show the exponential decay of nonlocal resistance with channel length. The inset is the calculated valley-relaxation length $L_v$ at different temperatures. The error bars are acquired from the linear fitting to $\ln \alpha_{NL} \sim L$. Since there is only three diffusion length in our study, we may underestimate the amount of error.

**Figure 6| Illustration of the MOKE experimental set-up and ROT-MOKE data of the Cd$_3$As$_2$ bulk crystals.** (**a**) Schematic drawing of the MOKE experimental set-up. The magnetic field is rotated in the y-z plane, parallel to the sample surface. A constant current density was applied across the sample. The incident laser is *p*-polarized with a wavelength of 670 nm. (**b** and **c**) ROT-MOKE signals of the Cd$_3$As$_2$ crystals under different current density and magnetic field, adopting a cosine-function dependence on *θ*. Here *θ* is defined as the angle between **E** and **B**. For (**b**) the magnetic field was fixed at 2000 Oe and for (**c**) the current density is fixed at 144 mA mm$^{-2}$. The insets are the data plotted in the polar coordinates.

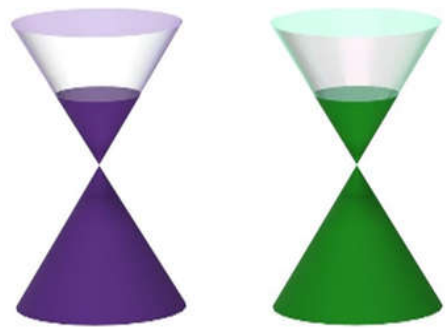
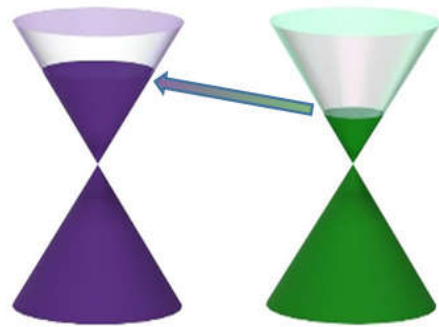
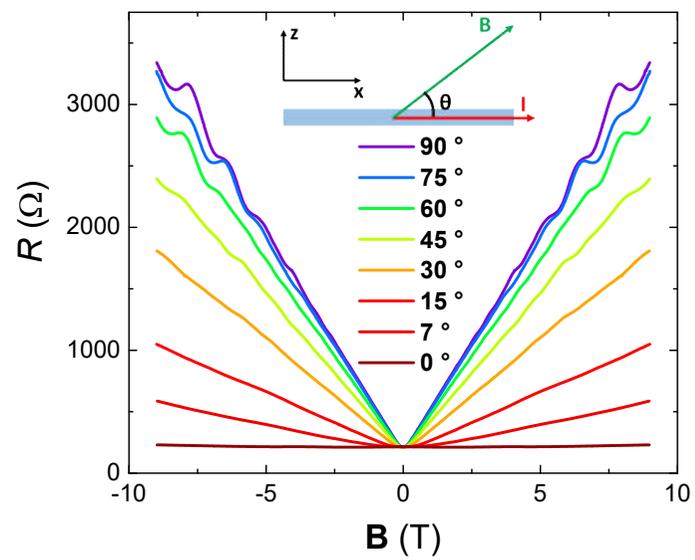
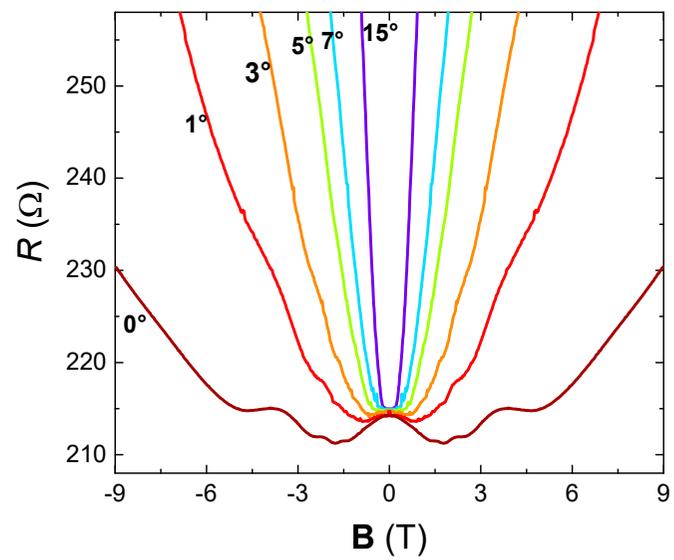

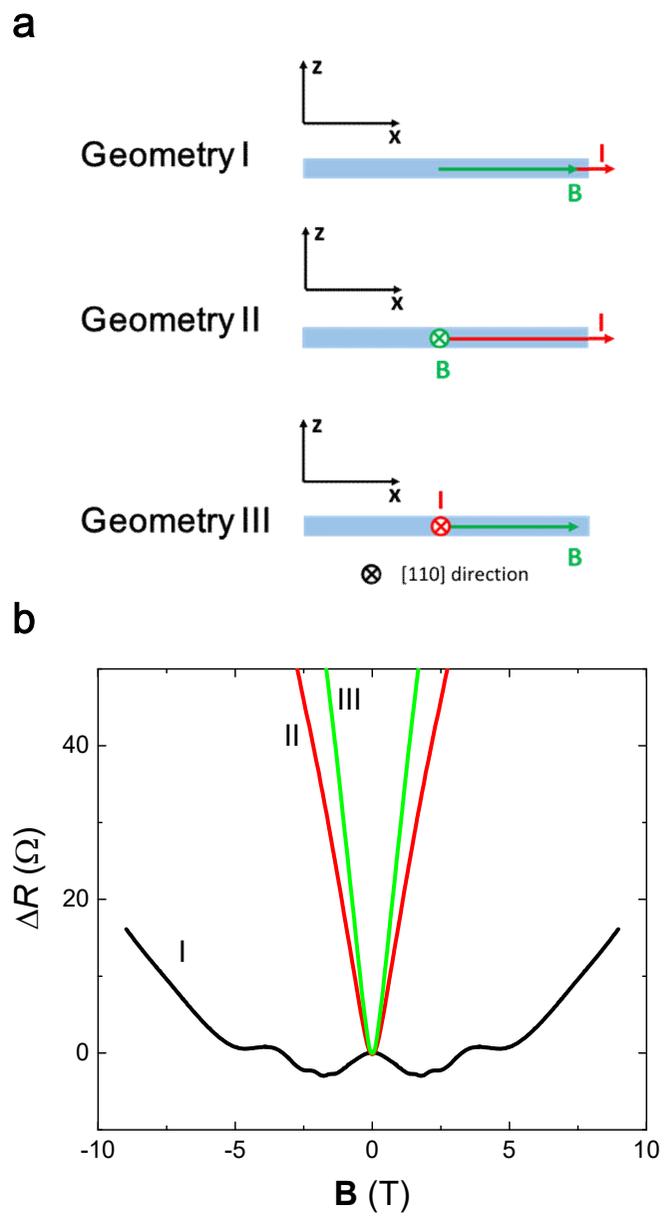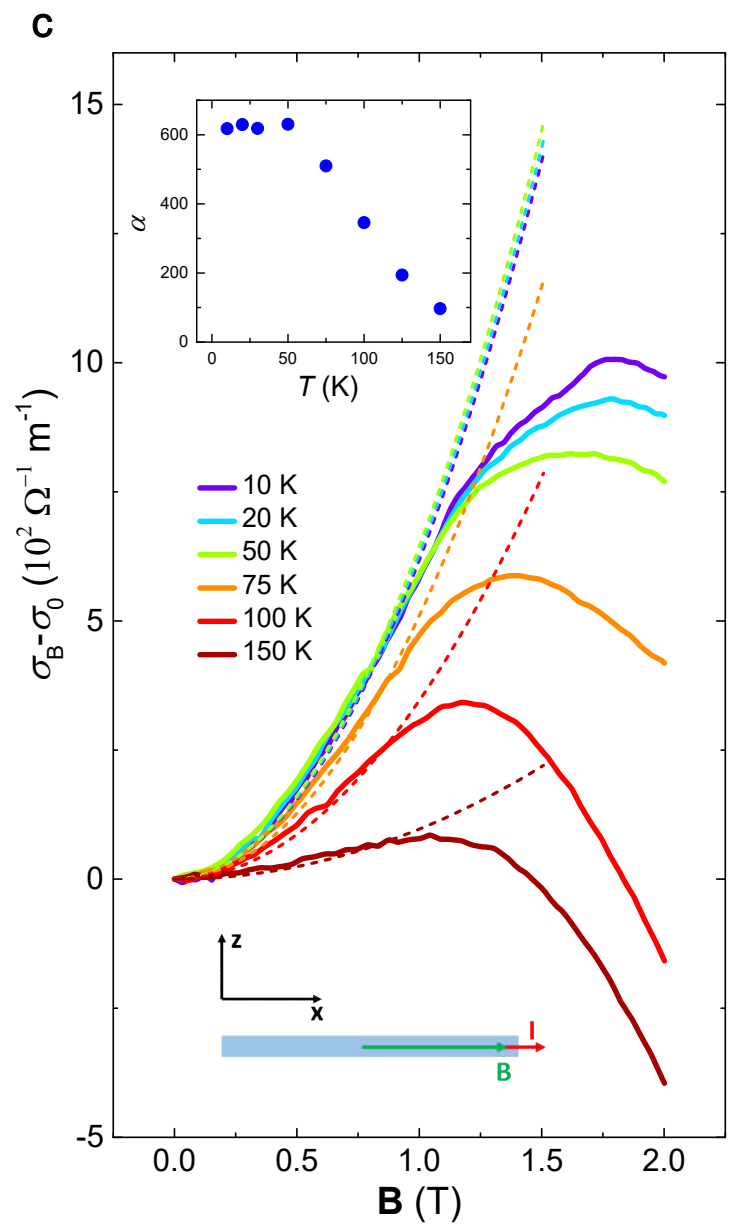

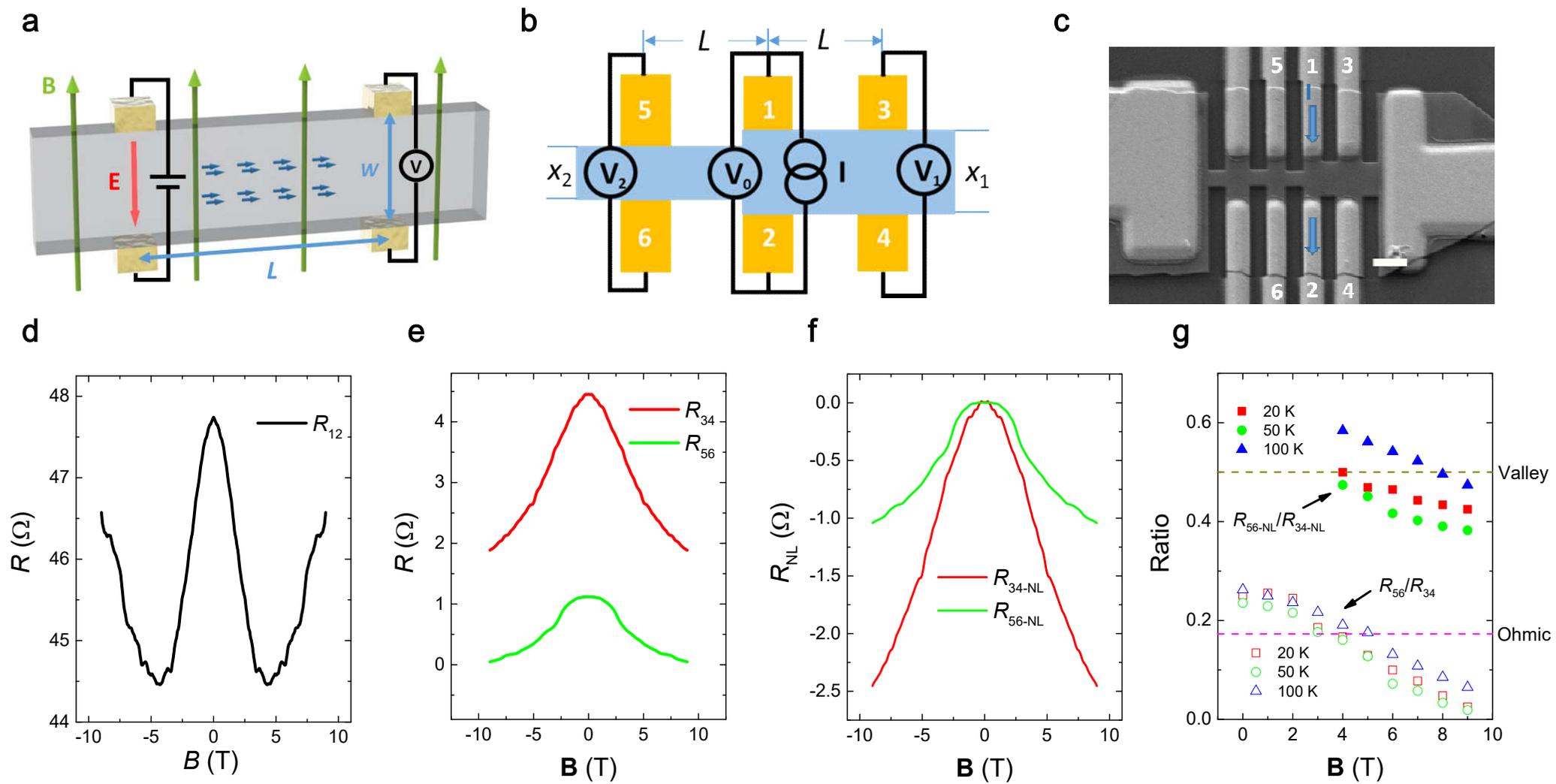

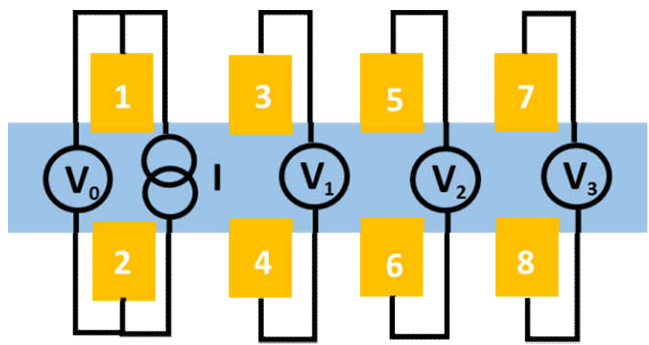
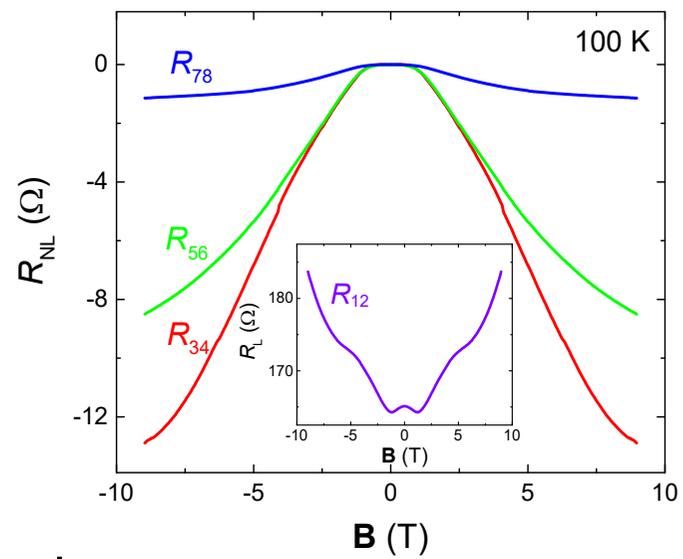
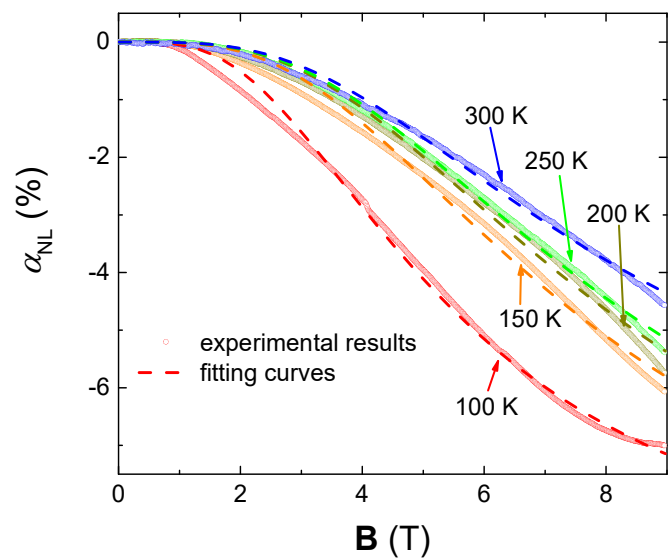
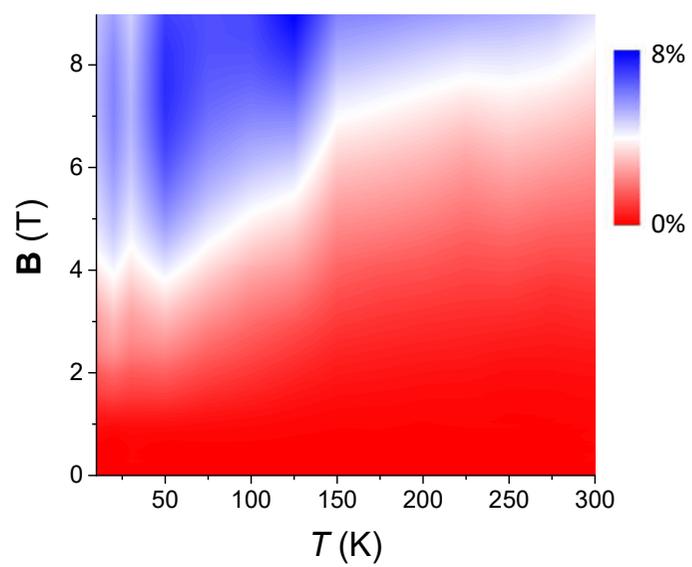

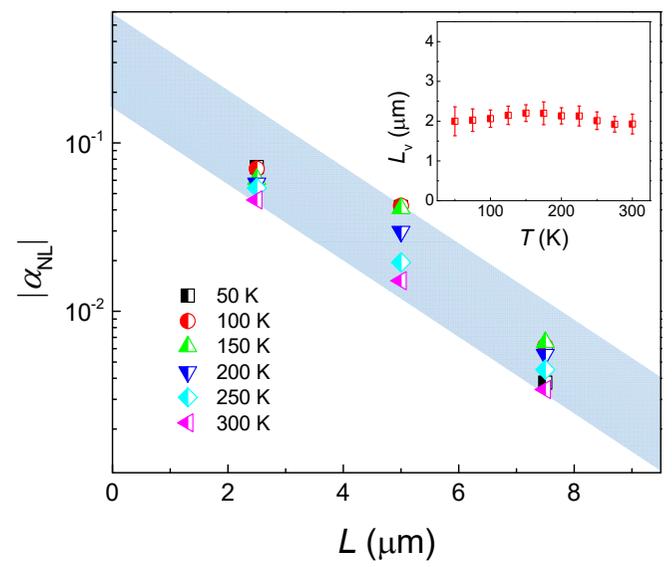

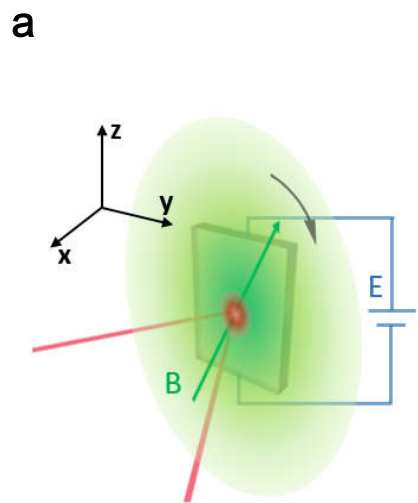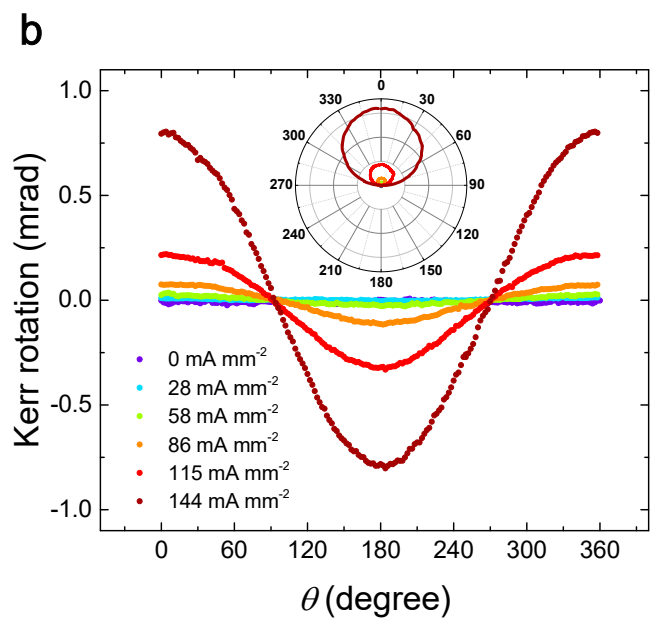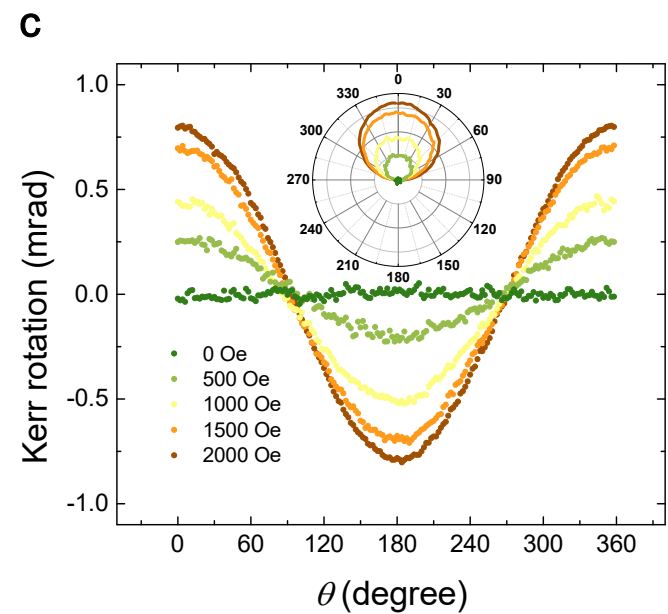